\documentclass[a4paper,12pt]{article}
\font\header=cmssdc10 at 20pt

\usepackage[ansinew]{inputenc}   
\usepackage{amsfonts}
\usepackage{indentfirst}
  \usepackage[normalem]{ulem}
  \usepackage{graphicx}
  \usepackage{amssymb}
 \usepackage{draftcopy}
  \usepackage{amsmath}
  \usepackage{rotating}
  \usepackage{color}
  \usepackage{setspace}
  \usepackage{fancybox}
 \usepackage{fancyvrb}
	\usepackage{fancyhdr}
	\usepackage[top=0.5cm,left=1cm,right=1cm,bottom=0.5cm]{geometry}
  \usepackage{wrapfig}
    \usepackage{float}
  \usepackage{type1cm}
\usepackage{eso-pic}
\usepackage{color}
 \usepackage{amsthm,amsmath,amsfonts}
\usepackage{fullpage,enumerate}
\usepackage{tikz}
\usepackage[colorlinks,citecolor=blue,urlcolor=blue]{hyperref}
\usepackage{hypernat}
\usepackage[notref,notcite]{showkeys}
\usepackage{graphicx}
\newcommand{\la}{\lambda}

\def\la{\lambda}

\def\1{{\bf 1}}

\begin{document}

{\header Can evolution paths be explained by chance alone?}

\vskip1cm

Rinaldo B. Schinazi

Department of Mathematics

University of Colorado at Colorado Springs

rinaldo.schinazi@uccs.edu

\vskip1cm

{\bf Abstract}  We propose a purely probabilistic model to explain the evolution path of a population maximum fitness. We show that after $n$ births in the population there 
are about $\ln n$ upwards jumps. This is true for any mutation probability and any fitness distribution and therefore suggests a general law for the number of upwards jumps. 
Simulations of our model show that a typical evolution path has first a steep rise followed by long plateaux. Moreover, independent runs show parallel paths. This is consistent with what was observed by Lenski and Travisano (1994) in their bacteria experiments.

\vskip1cm


\vskip1cm

{\header The model}

\vskip1cm

Lenski and Travisano (1994) studied the evolution of the bacteria Escherichia coli  for 10,000 generations. They graphed the average fitness as a function of time.
We were intrigued by the regularity of their curves (steep rise at first and then long plateaux) and the fact that different lines of bacteria had distinct but somehow parallel evolution paths.
Can a simple probability model account for such curves? This is what we attempt to answer in this paper.

We now describe our model. Let $\mu$ be the probability distribution for individual fitness. Let $0<s<1$ be the mutation probability per birth.
We start with a single individual whose fitness is a $\la_0$ sampled from the distribution $\mu$.  At every discrete time unit there is exactly one birth in the population.
There are two possibilities.

$\bullet$ With probability $1-s$ the
new individual has no mutation. We assign to the new individual a fitness that has previously appeared in the population. The exact way a fitness is assigned to the new individual is not important for our purposes. 
We could for instance sample a random  individual in the population and assign the sampled individual fitness to the new individual.

$\bullet$ With probability $s$ the new individual has a mutation. We sample a new $\lambda'$ from the distribution $\mu$. This new $\la'$ will be the
fitness of the new individual. 

Assuming mutations appear or not independently at each birth, the number of mutations at time $n$ 
has a binomial distribution $b(n,s)$ with parameters $n$ and $s$. Assuming the distribution $\mu$ is continuous every sampled  fitness
is distinct from all previously sampled fitnesses.  Hence, at time $n$ there is a sequence  $\la_0,\la_1,\la_2,\dots, \la_{b(n,s)}$ (numbered in the order of appearance) of distinct fitnesses that have
been sampled from the distribution $\mu$.  Moreover, every individual born up to time $n$ has been assigned one of these $b(n,s)$ fitnesses.

 \vskip1cm

{\header The number of upwards jumps}

\vskip1cm

Let $r(n)$ be the number of records in the sequence
 $\la_0,\la_1,\la_2,\dots, \la_{b(n,s)}$. 
 More precisely, $r(n)$ is the number of indices $k$ in $\{0,1,\dots,b(n,s)\}$ such that
 $$\la_k=\max\{\la_0,\la_1,\dots,\la_k\}.$$
 In words, $r(n)$ is the number of times that the sequence of fitnesses reaches a maximum.
 Hence, at time $n$ there are $r(n)$ upwards jumps in the evolution path.

Assume that $\mu$ is a continuous probability distribution with support in $(0,+\infty)$. 
Our main result is
$$\lim_{n\to\infty} \frac{r(n)}{\ln n}=1.\leqno(1)$$

Here are a few consequences of this result.

$\bullet$ The limit in (1) does not depend on the mutation probability $s$ nor on the probability distribution $\mu$! That is, in this model  the number of upwards mutations follows a general law which is independent of the mutation probability and of the distribution of the fitness.

$\bullet$ By the Law Large Numbers the total number of mutations $b(n,s)$ up to time $n$ is of order $ns$  for $n$ large. Hence, (1) shows that the fraction of upwards mutations as compared to the total number of mutations is of order $\frac{\ln n}{ns}$.

$\bullet$  When an upwards mutation appears it need not persist in the population. By bad luck it may disappear before giving birth. Hence, $\ln n$ is really an upper bound on the number of upwards mutations.

\vskip1cm

{\header Simulation}

\vskip1cm

We graph the maximum fitness as a function of time for 
 three independent runs, see the figure below. 
We use for $\mu$ (i.e. the fitness of the distribution) a mean 1 exponential distribution and we take the
mutation probability per birth to be $s=0.02$. For each run the total number of births is $10^6$. The initial individual is sampled independently for each run from $\mu$. The number of jumps per run
are 11, 11 and 7, respectively.
We observe a steep rise in the beginning followed by long plateaux.
The three runs are somewhat parallel.


\vskip1cm

{\header Literature}

\vskip1cm

There is a vast literature on adaptive walks that goes back to at least Gillepsie (1983), see also Orr (2002) and (2003). Our model is somewhat related to these models. However, our point of view is different. Many papers are concerned with the number of steps that a single DNA sequence makes in a certain fitness landscape before being stuck at a local maximum. On the other hand we are concerned with the global maximum of a population for which all jumps are allowed. Our main purpose is to test whether chance alone can produce an evolution path for the whole population consistent with biological observations.

Next we describe an adaptive walk first studied by Kauffman and Levin (1987) for which there is a result reminiscent of ours.  In that model a genome is a string of $n$ sites. Every site has a $0$ or a $1$. The genome evolves by flipping one site at a time. Each of the $2^n$ possible configurations of the genome is assigned a fitness a priori. In the simplest variation of this model fitnesses are assigned independently to every configuration according to a uniform distribution. An "accessible path" is a path going from all $0's$ configuration to all $1's$ configuration along a sequence of configurations that increases fitness at every step. Hegarty and Martinsson (2014) proved that the probability that an accessible path exists is of order $\frac{\ln n}{n}$ as $n$ goes to infinity. Note that the fraction of upwards mutations in our model is of the same order.

\vskip1cm

{\header Discussion}

\vskip1cm

Our model shows that by chance alone one can obtain evolution paths consistent with those obtained experimentally by Lenski and Travisano (1994) for bacteria lines. Moreover, independent runs of our model have somewhat parallel trajectories, see the figure below. This is also what is observed by Lenski and Travisano (1994) for different bacteria lines. They formulate two hypotheses to explain this.  The first is that the bacteria lines evolve in identical environments and the second is that large population sizes will trigger identical mutations in both lines. In our model the environment is determined by the fitness distribution $\mu$. We also find that the specific trajectories depend on $\mu$ but the general shape of the trajectory (i.e. fast rising at first and then reaching successive plateaux) does not. As for the second hypothesis, in our model two mutations are never identical and we still get parallel trajectories.   Hence, our model gives arguments against the hypotheses formulated by 
Lenski and Travisano (1994) but supports their general conclusion that  "our experiment demonstrates the crucial role of chance events".

\vskip1cm

{\header Proof of (1)}

\vskip1cm

The proof of (1) is a consequence of a classical result in probability theory that we now state. Let $X_1,X_2,\dots,X_n$ be a sequence of independent identically distributed random variables with the same continuous distribution. Let $R_n$ be the number of records in this sequence. That is, $R_n$ is the number indices $k$ in $\{1,2,\dots,n\}$ such that
$$X_k=\max\{X_1,X_2,\dots,X_k\}.$$
Then,
$$\lim_{n\to\infty} \frac{R_n}{\ln n}=1.\leqno (2)$$
For a proof of (2) see for instance Port (1994).

We now turn to the proof of (1).
Recall that the total number of mutations up to time $n$ is $b(n,s)$, a binomial random variable with parameters $n$ and $s$.
Hence, the sequence of $\la$'s up to time $n$ has length $b(n,s)$ and $r(n)$ (i.e. the number of records in that sequence) is such that $r(n)=R_{b(n,s)}$.
Since $b(n,s)$ goes to infinity with $n$ (by the Law of Large Numbers) we have by (2) that
$$\lim_{n\to\infty} \frac{r(n)}{\ln b(n,s)}=1,$$
where the convergence above as well as all subsequent convergences occurs almost surely (i.e. with probability one).
By the Law of Large Numbers
$\lim_{n\to\infty} \frac{b(n,s)}{n}=s>0$. Therefore,  $\lim_{n\to\infty}\frac{\ln b(n,s)}{\ln n}=1.$

Observe that
$$\frac{r(n)}{\ln n}=\frac{r(n)}{\ln b(n,s)}\frac {\ln b(n,s)}{\ln n}.$$
We have shown that both fractions on the r.h.s. converge to 1. So does the l.h.s. The proof of (1) is complete.

\bigskip

{\bf Remark.}  Intuitively one may expect that a higher probability mutation will trigger more upwards mutations. Our result (1) shows that in the long run this is not so.
The total number of mutations is increasing as a function of $s$ but the number of upwards mutations is not. We explain why.
As noted before the total number of mutations $b(n,s)$ up to time $n$ has a binomial distribution with parameters $n$ and $s$. Let $0<s_1<s_2<1$. 
It is easy to couple the models with mutation probabilities $s_1$ and $s_2$ respectively so that for every $n\geq 1$, $b(n,s_1)\leq b(n,s_2)$ and
$$\{\la_0,\dots, \la_{b(n,s_1)}\}\subset \{\la_0,\dots, \la_{b(n,s_2)}\}.$$
In particular, at any given time the model with $s_2$ has had more mutations and has a larger maximum fitness than the model with $s_1$. 
On the other hand the model with $s_2$ need not have more upwards mutations than the model with $s_1$. 

\vskip1cm

{\header References}

\vskip1cm

J.H. Gillepsie (1983) A simple stochastic gene substitution model. Theoretical  Population Biology 23, 202-215.

P. Hegarty and S. Martinsson (2014) On the existence of accessible paths in various models of fitness landscapes. Annals of Applied Probability 24, 1375-1395.

S. Kaufman and S. Levin (1987) Towards a general theory of adaptive walks in rugged landscapes. Journal of theoretical biology 128, 11-45.

R.E. Lenski and M. Travisano (1994) Dynamics of adaptation and diversification: a 10,000-generation experiment with bacterial populations. PNAS 91, 6808-6814.

H.A. Orr (2002) The population genetics of adaptation: the adaptation of DNA sequences. Evolution 56, 1317-1330.

H.A. Orr (2003) A minimum on the mean number of steps taken in adaptive walks. Journal of theoretical biology 220, 241-247.

S. C. Port (1994) {\it Theoretical probability for applications}. Wiley.

\begin{figure}[ht]
\includegraphics[width=15cm]{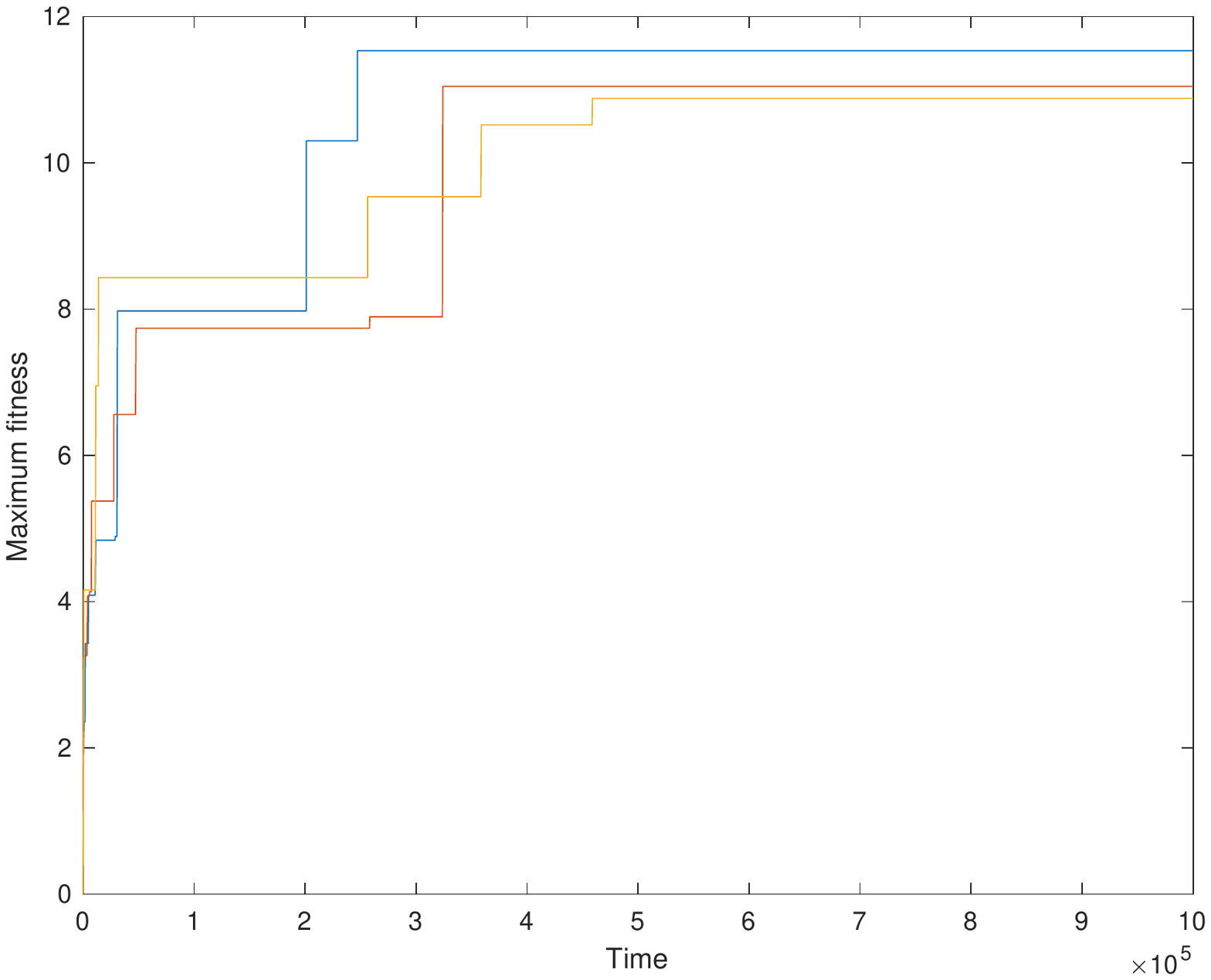}
\end{figure}

\end{document}